\documentclass[aps,prd,10pt,onecolumn,nofootinbib,showpacs,showkeys,notitlepage]{revtex4-1}
\usepackage{amsmath,graphicx,bm}
\usepackage{hyperref}

\renewcommand{\d}{\partial}

\newcommand{\nn}{\nonumber\\}

\newcommand{\rh}{\varrho}

\newcommand{\exv}[1]{\left\langle{#1}\right\rangle}

\newcommand{\ep}{\varepsilon}

\renewcommand{\k}{{\bf k}}

\newcommand{\Disc}{\mathop{\textrm{Disc}}}
\renewcommand{\Im}{\,\textrm{Im}\,}
\renewcommand{\Re}{\,\textrm{Re}\,}
\newcommand{\pint}[2]{{\int\!\frac{d^{#1}#2}{(2\pi)^#1}\,}}

\renewcommand{\v}{\nu}

\newlength{\szovszel}\newlength{\szovmag}

\newcommand{\MSbar}{\ensuremath{\overline{\mathrm{MS}}}}
\newcommand\lsim{\mathrel{\rlap{\lower4pt\hbox{\hskip1pt$\sim$}} \raise1pt\hbox{$<$}}}                
\newcommand\gsim{\mathrel{\rlap{\lower4pt\hbox{\hskip1pt$\sim$}} \raise1pt\hbox{$>$}}}                

\newcommand{\tehat}{{\ensuremath\quad\Rightarrow\quad}}

\begin{document}

\title{Resummations in the Bloch-Nordsieck model}

\author{A. Jakov\'ac}
\email{jakovac@phy.bme.hu}
\author{P. Mati}
\email{mati@phy.bme.hu}
\affiliation{Institute of Physics, Budapest University of Technology
  and Economics, H-1111 Budapest, Hungary}




\begin{abstract}
  We studied different levels of resummations of the exactly solvable
  Bloch-Nordsieck model in order to be able to compare the
  approximations with an exact result. We studied one-loop
  perturbation theory, 2PI resummation and Schwinger-Dyson equations
  truncated in a way to maintain Ward-identities. At all levels we
  carefully performed renormalization. We found that although the 2PI
  resummation does not exhibit infrared (IR) sensitivity at the mass
  shell, as the one-loop perturbation theory does, but it is still far
  from the exact solution. The method of truncated Schwinger-Dyson
  equations, however, is exact in this model, so it provides a new
  way of solving the Bloch-Nordsieck model. This method can also be
  generalized to other, more complicated theories.
\end{abstract}

\maketitle

\section{Introduction}
\label{sec:intro}

In field theories we often encounter infrared (IR) divergences of
different kind. Some of them have physical meaning while others are
just artifacts of the perturbation theory. A common feature, however,
is that the IR divergences appear at each order of perturbation
theory, usually being more and more severe at higher loop orders. In
order to consistently define the theory, these IR divergences must be
summed up.

2PI resummations provide a consistent resummation framework known for
a long time \cite{2PIhist}. The basic idea is to replace the free
propagator in the perturbation theory with the exact one which is
approximated self-consistently with fixed-loop skeleton diagrams. The
so-defined perturbation theory is renormalizable
\cite{HeesKnoll}-\cite{Fejos:2009dm}, and can be applied to study
different physical questions from non-equilibrium \cite{BergesCox},
\cite{BBW}, thermodynamics \cite{Jakovac},
\cite{Berges:2004hn},\cite{BR},\cite{BRo} and different systems like
O(N) model \cite{Jakovac1}, \cite{Fejos:2009dm} or gauge theories
\cite{RS1}.

Although the 2PI approximation is constructed by physical arguments,
and we expect better results (ie. closer to the exact one) after 2PI
resummation, a priory it is not sure that one really achieves this
goal. Probably the finite lifetime effects are well represented by 2PI
resummation both in equilibrium \cite{Jakovac} as well in
non-equilibrium, where the 2PI is close to the Boltzmann-equation
approximation \cite{CK}. But if the deep IR regime is important where
multi-particle excitations also play crucial role, the picture is far
to be so clean. To make the case even worse, in most gauge theory
models there is hard to make exact statements about the IR behavior
of the model.

In this paper we aim to study the Bloch-Nordsieck model, which is an
exactly solvable 3+1D gauge theory \cite{BogoljubovShirkov}. It is the
eikonal approximation model of the QED, and one can argue
\cite{IancuBlaizot} that in the deep IR regime it describes correctly
QED. Therefore it is used to estimate IR properties of QED from this
model, for example the soft photon creation processes \cite{Weldon} or
finite temperature electron damping \cite{IancuBlaizot}.

This model is therefore a motivated case study where the accuracy of
the 2PI resummation can be assessed. We therefore perform a series of
approximations for the Bloch-Nordsieck model: a one-loop perturbation
theory, a 2PI resummation and finally the solution of the
Schwinger-Dyson equations with an Ansatz for the vertex function. In
this model all levels of the approximations can be treated
analytically. We show that the last method is exact in the model under
consideration -- although that is not expected in general. This
observation, however, leads us to a proposal how the 2PI resummation
can be improved in a generic model in order to catch the IR physics
correctly.

The structure of the paper is as follows. We first overview the
standard solution of the Bloch-Nordsieck propagator at zero
temperature in Section \ref{sec:1}. Then we compute the one loop level
fermion propagator in Section \ref{sec:oneloop}. Next, in Section
\ref{sec:2PI} we write up the 2PI equations for this model, perform
renormalization on that, and give the analytical solution as far it
can be done. Then we study the problem also numerically, determine the
fermion propagator and compare the result with the exact one. Finally,
in Section \ref{sec:SD} we study the truncated Schwinger-Dyson
equations, renormalize it, and show that for the Bloch-Nordsieck model
the so-defined approximation scheme is exact. For conclusion (Section
\ref{sec:conc}) we summarize our approach again and speculate about
the possible generalizations.

\section{The model and the exact solution}
\label{sec:1}

The Bloch-Nordsieck model is a simplification of the QED Lagrangian
where the Dirac matrices $\gamma^\mu$ are replaced by a four-vector
$u^\mu$
\begin{equation}
  {\cal L} = -\frac14 F_{\mu\nu} F^{\mu\nu} + \bar \Psi(iu_\mu D^\mu
  - m)\Psi,\qquad iD_\mu = i\d_\mu -eA_\mu,\quad F_{\mu\nu}=\d_\mu
  A_\nu - \d_\nu A_\mu.
\end{equation}
The singled-out four vector $u_\mu$ represents the velocity of the
rest frame of the fermion. The fermion wave function here has only one
component and $\bar \Psi=\Psi^*$.

We are interested in the fermion propagator which reads in the path
integral representation
\begin{equation}
  i{\cal G}(x)=\exv{T \Psi(x)\bar\Psi(0)} = \frac1Z \int{\cal D}\bar\Psi
  {\cal D}\Psi {\cal D}A_\mu e^{iS[\bar\Psi,\Psi,A]}\Psi(x)\bar\Psi(0).
\end{equation}
At the tree level it reads:
\begin{equation}
  {\cal G}_0(p) = \frac1{u_\mu p^\mu -m +i\ep}.
\end{equation}
Since it has a single pole, there is no antiparticles in the model,
and also the Feynman propagator is the same as the retarded
propagator. The lack of antiparticles also means that all closed
fermion loops are zero\footnote{This statement can be best seen in
  real time representation. There a chain of fermion propagators,
  because of the retardation, is proportional to $\Theta(t_1-t_2)\dots
  \Theta(t_{n-1}-t_n)$. In a closed loop $t_n=t_1$, therefore the
  product of theta functions is zero almost everywhere.}.  As a
consequence the photon self-energy is zero, the free photon propagator
is the exact one. In Feynman gauge therefore the \emph{exact} photon
propagator is
\begin{equation}
  G_{\mu\nu}(k) = \frac{-g_{\mu\nu}}{k^2+i\ep}.
\end{equation}

Now we shortly review the exact solution for the fermion propagator,
cf. \cite{BogoljubovShirkov}, \cite{Fried}. We first define the gauge
field dependent propagator:
\begin{equation}
  {\cal G}(x;A) = \int{\cal D}\bar\Psi {\cal D}\Psi
  e^{iS[\bar\Psi,\Psi,A]}\Psi(x)\bar\Psi(0).
\end{equation}
This satisfies the Schwinger-Dyson equation
\begin{equation}
  (iu_\mu\d^\mu-eu_\mu A^\mu - m) {\cal G}(x;A) = -\delta(x).
\end{equation}
We will need the solution in an exponential form for $A$, and this is
achieved by introducing the auxiliary quantity $U(x,\nu)$ which satisfies
\begin{equation}
  \label{Ueq}
  i\frac{\d U(x,\nu)}{\d\nu} = (iu_\mu\d^\mu-eu_\mu A^\mu - m)
  U(x,\nu),\qquad U(x,0)=\delta(x). 
\end{equation}
By integration of the above equation by $\nu$ and assuming
$U(x,\infty)=0$ (for which we need a convergence factor $i\ep$) we see
that
\begin{equation}
  {\cal G}(x)=  -i\int d\nu U(x,\nu).
\end{equation}
We perform Fourier transformation and separate the free time
dependence $U(p,\nu) = e^{-i(u_\mu p^\mu - m)\nu} \bar
U(p,\nu)$, then we obtain
\begin{equation}
  \frac{\d\bar U(p,\nu)}{\d\nu} = ieu_\mu \pint4k A^\mu(p-k)
  e^{iu^\mu(p_\mu-k_\mu)} \bar U(k,\nu),\qquad \bar U(p,0)=1.
\end{equation}
The linear $p_\mu-k_\mu$ behavior is the consequence of the linearity
in the kernel. If the kernel is non-linear or it is not scalar (has a
matrix structure) then this form is not true any more. From here an
inverse Fourier transformation yields
\begin{equation}
  \frac{\d\bar U(x,\nu)}{\d\nu} = ieu_\mu A^\mu(x+u\nu) \bar U(x,\nu)
  \qquad\Rightarrow \qquad \bar U(x,\nu) = e^{\int\limits_0^\nu d\nu'
    A(u\nu')} \delta(x).
\end{equation}
Once we have an exponential representation for the
background-dependent propagator, we can perform the Gaussian
$A$-integration. As a result we obtain in the exponent the factor
\begin{equation}
  \frac {ie^2}2 \pint4k R_\nu^*(k)G_{\mu\nu}(k) R_\nu(k),\qquad R_\nu(k) =
  \int\limits_0^\nu d\nu' e^{-ik_\mu u^\mu \nu'}.
\end{equation}
This integral is UV divergent; in dimensional regularization one finds
the result
\begin{equation}
  \frac{\alpha}{2\pi\ep} + \frac\alpha\pi \ln i\bar \mu\nu,
\end{equation}
where $\alpha=e^2/(4\pi)$ and $\bar\mu =
\sqrt{4\pi}e^{\gamma_E}\mu$. Then the fermion propagator reads
\begin{equation}
  {\cal G}(p) = -i e^{\frac{\alpha}{2\pi\ep}} \int\limits_0^\infty d\nu\,
  e^{-i\nu (u_\mu p^\nu-m) + \frac\alpha\pi \ln i\bar\mu\nu } = \frac
  {Z}{(u_\mu p^\nu-m)^{1+\frac\alpha\pi}}
\end{equation}
where $Z=\Gamma(1+\frac\alpha\pi) e^{\frac{\alpha}{2\pi\ep}}
{\bar\mu}^{\frac\alpha\pi}$. This is UV divergent which means that we
need a wave function renormalization. The renormalized propagator
reads
\begin{equation}
  \label{Gex}
   {\cal G}_{ren}(p) = \frac \zeta{(u_\mu p^\nu-m)^{1+\frac\alpha\pi}},
\end{equation}
where $\zeta$ is a finite quantity.

We can determine the discontinuity of this formula, for simplicity
choosing $u_\mu=(1,0,0,0)$:
\begin{equation}
  \rh(p) = \Disc_{p_0} {\cal G}(p) = \Theta(p_0-m)\,
  \frac{\zeta(1-e^{2i\alpha})}{(p_0-m)^{1+\frac\alpha\pi}}.
\end{equation}
With this spectral function the sum rule $\int\limits_{-\infty}^\infty
dp_0 \rh(p_0) =1$, which is the consequence of the equal time
anticommutation relations, cannot be fulfilled, since the integral is
divergent. This divergence should be compensated with the choice
$\zeta=0$, but then we are faced with a $0\times \infty$
expression. Therefore one should always use a regularized version of
the spectral function (or propagator), maintaining the sum rule, and
only at the end of the calculation is one allowed to release the
regularization.

The Lagrangian is Lorentz-invariant in the sense that we must also
transform $u$. So we can choose a Lorentz-transformation where
$\Lambda u=(u_0,0,0,0)$. If $u^\mu$ is a 4-velocity then $u_0=1$; if
it is of the form $u=(1,\mathbf{v})$, then it is
$u_0=\sqrt{1-\mathbf{v}^2}$. After rescaling the field $\Psi\to
\Psi/\sqrt{u_0}$ and the mass as $m\to u_0m$, the Lagrangian reads
\begin{equation}
  {\cal L} = -\frac14 F_{\mu\nu} F^{\mu\nu} + \bar \Psi(iD_0-m)\Psi.
\end{equation}
This Lagrangian will be used mostly in this work later. If necessary,
the complete $u$ dependence is easily recoverable.

\section{One loop perturbation theory}
\label{sec:oneloop}

The goal of our investigations is to see, how the different levels of
resummations improve the result. Thus first we start with the one loop
perturbation theory. Here we need the renormalized Lagrangian; in
Feynman gauge it reads (using the fact that the photon self-energy is
zero):
\begin{equation}
  {\cal L} = -\frac12 (\d_\mu A_\nu)^2 + \bar \Psi (i\d_0 - m)\Psi -
  e\bar\Psi A_0\Psi + \delta Z\bar \Psi i\d_0\Psi  - \delta Z_m m \bar\Psi
  \Psi - \delta e\bar\Psi A_0\Psi.
\end{equation}
For the fermion self-energy the one loop diagram is the bubble with
the contribution:
\begin{equation}
  \label{oneloop}
  -i\Sigma_{1loop}(p,m) = (-ie)^2 \pint4k iG_{00}(k)\, i{\cal
    G}(p-k)= -e^2u^2 \pint4k \frac1{k^2+i\ep} \frac1{p_0-k_0-m+i\ep}.
\end{equation}
Moreover we have wave function and mass renormalization counterterms
\begin{equation}
  \Sigma_{ct}(p) = -\delta Z p_0 + \delta Z_m m.
\end{equation}
The complete one loop self-energy is $\Sigma_{1loop}+\Sigma_{ct}$. In
the calculation we have to take care of the non-standard form of the
free fermion propagator. The details of the computation can be found
in the Appendix, as a result we obtain
\begin{equation}
  \label{Sigma1}
  \Sigma_{1loop}(p,m) = \frac{\alpha}{\pi}(p_0-m) \left[
    -\ln\frac{m-p_0}\mu + {\cal D}_\ep\right],
\end{equation}
where $\alpha=e^2/(4\pi)$ and
\begin{equation}
  {\cal D}_\ep =\frac1{2\ep}+1+\frac12(\ln\pi-\gamma_E).
\end{equation}
For renormalization we have to subtract the divergences with help of
the counterterms, the finite parts are fixed by the renormalization
scheme. In the \MSbar\ scheme we choose the counterterms like
\begin{equation}
  \delta Z_{1,\MSbar} = \delta Z_{m,\MSbar} = \frac{\alpha}{\pi}{\cal D}_\ep,
\end{equation} 
this results in
\begin{equation}
  \label{Sigmaren1}
  \Sigma_{ren}(p) = -\frac{\alpha}{\pi}(p_0-m)\ln\frac{m-p_0}\mu.
\end{equation}

The discontinuity of the renormalized self-energy reads
\begin{equation}
  \label{Disc1}
  \Disc_{p_0} \Sigma(p) = 2\alpha\,(p_0-m) \Theta(p_0-m).
\end{equation}
For the one-loop propagator we obtain
\begin{equation}
  {\cal G}(p) = \frac1{p_0-m -\Sigma(p)} =
  \frac1{p_0-m}\,\frac1{1+\displaystyle
    \frac{\alpha}{\pi}\ln\frac{m-p_0}\mu}.
\end{equation}
This is consistent with the exact result \eqref{Gex} in the leading
order of $e^2$.

The spectral function $\Disc_{p_0}i{\cal G}$ reads
\begin{equation}
  \rh(p) = \frac{\Theta(p_0-m)}{p_0-m}\,\frac{2\alpha} {\displaystyle
    \left(1+ \frac\alpha\pi\ln\frac{p_0-m}\mu\right)^2 + \alpha^2}.
\end{equation}
This spectral function is normalizable, since
\begin{equation}
  \int\limits_{-\infty}^\infty \frac{dp_0}{2\pi} \rh(p) = \frac\pi\alpha.
\end{equation}

On the other hand the one-loop result is not reliable when
$|\ln(p_0-m)/\mu|\gg \frac\pi\alpha$, ie. in the vicinity of the mass
shell as well as in the large $p_0$ regime. In order to have a better
description of these kinematical regimes, we need resummation of certain
class of diagrams.

\section{2PI resummation}
\label{sec:2PI}

As it is discussed in the Introduction, the next level of our
approximations is the 2PI resummation. The idea is to use the exact
propagators in the perturbation theory, this propagator is determined
self-consistently using skeleton diagrams as resummation patterns. The
one-loop bubble diagram in the present case generates the resummation
of all the ``rainbow'' diagrams. To obtain an expression for the 2PI
resummation we use the technique of \cite{Jakovac}: we use the 1loop
formula \eqref{oneloop}, interpret the appearing propagators as full
propagators, and finally perform renormalization with the same
\emph{form} of divergent parts of the counterterms as in the 1-loop
case (the actual values will be different). 

The tree level photon propagator is exact, therefore we can write
\begin{equation}
  \Sigma(p) = -ie^2 \pint4k \frac{{\cal G}(p-k)}{k^2+i\ep}.
\end{equation}
Using a spectral representation for the fermion propagator (using that
now the Feynman propagator is the retarded one and that the fermion
spectral function is $\rh(\omega<0)=0$) we find
\begin{equation}
  \Sigma(p) = -ie^2 \int\limits_0^\infty \frac{d\omega}{2\pi}
  \pint4k \frac1{k^2+i\ep} \frac{\rh(\omega)}{p_0-k_0-\omega+i\ep}.
\end{equation}
From this form it is clear that we obtain the weighted one-loop
result, ie.
\begin{equation}
  \label{2PIsprep}
  \Sigma(p) =  \int\limits_0^\infty \frac{d\omega}{2\pi} \rh(\omega)
  \Sigma_{1loop}(p,\omega).
\end{equation}
In particular, if $\rh(\omega)=2\pi\delta(\omega-m)$, then we get back
the one-loop result. 

At this point it is worth to examine the UV divergence structure of
the 2PI approximation. UV divergences may occur in \eqref{2PIsprep}
for large values of $\omega$: using \eqref{Sigma1} we find that the
large $\omega$ behavior of the one-loop self-energy reads:
\begin{equation}
  \label{2PIdiv}
  \Sigma_{1loop}(p,\omega) = \frac\alpha\pi\omega \left(\ln\frac\omega\mu 
  -{\cal D}_\ep\right) + \frac\alpha\pi\left(-\ln\frac\omega\mu
    +{\cal D}_\ep\right) p_0 + {\cal O}(\frac{p_0^2}\omega).
\end{equation}
Since $\rh$ is integrable for large $\omega$ values, therefore the
${\cal O}(\omega^{-1})$ is already finite. Therefore the divergence
structure of the self-energy is $A+Bp_0$, just like for the free case,
and so the same type of counterterms are needed (although the values
are different). This is a manifestation of the general case of
counterterm renormalizability of 2PI resummations \cite{Jakovac}.

\subsection{Analytic study of the 2PI equations}
\label{sec:ex2PI}

First we try to analyze \eqref{2PIsprep} with analytic methods. We
differentiate it with respect to $p_0$ to find
\begin{equation}
  \frac{\d\Sigma_{1loop}}{\d p_0} = \frac\alpha\pi\left(
    -\ln\frac{\omega-p_0-i\ep}\mu -1\right),\qquad
  \frac{\d^2\Sigma_{1loop}}{\d p_0^2} = -\frac\alpha\pi
  \frac1{p_0-\omega+i\ep}, \qquad  \frac{\d^2\Sigma}{\d p_0^2} =
  -\frac\alpha\pi {\cal G}.
\end{equation}
Since ${\cal G}^{-1} = p_0-m-\Sigma$, we find for ${\cal G}^{-1}$:
\begin{equation}
  \frac{d^2{\cal G}^{-1}}{dp_0^2}{\cal G}^{-1} = \frac\alpha\pi.
\end{equation}
To solve the equation we first should realize that the $\alpha=0$ and
$\alpha\neq0$ cases are very different. If $\alpha=0$ then $({\cal
  G}^{-1})''=0$ and the propagator behaves as ${\cal G}= Z/(p_0 -
\tilde m)$ with some wave function renormalization constant $Z$ and
mass $\tilde m$. This agrees with the free case. We also see that the
integration constants correspond to the renormalization scheme (here
the wave function and mass renormalization).

If $\alpha\neq 0$ then we can redefine the variables with an arbitrary
${\cal G}_0$ scale as
\begin{equation}
  \label{EPsidef}
  E = {\cal G}_0 \sqrt{\frac{2\alpha}\pi}\, (m-p_0),\qquad \Psi = -
  {\cal G}_0 {\cal G}^{-1},
\end{equation}
then we find
\begin{equation}
  2\frac{d^2\Psi}{dE^2}\Psi=1.
\end{equation}
This equation does not depend on the coupling any more. The coupling
constant dependence shows up in the integration constants which are
the manifestation of the renormalization scheme. We shall also note
that the equation does not give information about the sign of $E$ and
$\Psi$, because for $E\to-E$ or $\Psi\to-\Psi$ the equation remains the
same. The chosen signs in \eqref{EPsidef} turn out later to be the
physical choice.

We introduce
\begin{equation}
  y = \frac{d\Psi}{dE} \tehat \frac{dy}{dE} = \frac{dy}{d\Psi}
  \frac{d\Psi}{dE} = E \frac{dy}{d\Psi}. 
\end{equation}
This means that we can write for $y$:
\begin{equation}
  2y\Psi\,  \frac{dy}{d\Psi} = 1 \tehat y = \frac{d\Psi}{dE}
  =\sqrt{\ln\Psi} + y_0,
\end{equation}
with an integration constant $y_0$. Therefore
\begin{equation}
  \label{sol}
  \int\limits_1^\Psi \! \frac{d\Psi'}{\displaystyle
    \sqrt{\ln \Psi'}+y_0} = E.
\end{equation}
There could appear an integration constant also here on the right hand
side: $E-E_0$. But recalling that $E\sim p_0-m$, we see that $E_0$
corresponds to a mass shift: if the mass remains the tree level $m$
then $E_0=0$.

This is the (implicit) solution of the 2PI equations. We see that for
real $\Psi$ the left hand side is real and positive, moreover for
$\Psi(E=0)=1$. The $E<0$ part corresponds to imaginary values of
$\Psi$. Since the equation itself is real, if $\Psi$ is a solution, it
is $\Psi^*$, too. This means that the imaginary part is in fact the
(half) discontinuity of the solution.

We see that irrespective of the value of $y_0$, at $E=0$, ie. on the
mass shell $\Psi=1$ and so ${\cal G}=-{\cal G}_0$ finite. This yields
difficulties when we try to apply renormalization conditions on the
self-energy. Namely, if we keep the mass shell unchanged (this would
correspond to the choice of $E_0$ above), then the renormalization of
the self-energy would mean $\Sigma(p_0=m)=0$ and $\Sigma'(p_0=m)=$
finite. Then, however, near the mass shell the propagator should
always behave as $\sim 1/(p_0-m)$, ie. \emph{infinite} at the mass
shell. This means that the physical renormalization process requires
${\cal G}_0\to\infty$. In this case the propagator behaves near the
mass shell as:
\begin{equation}
  {\cal G} = \frac{-{\cal G}_0}{1+\displaystyle {\cal G}_0y_0
    \sqrt{\frac{2\alpha}\pi} (m-p_0)}
  \stackrel{{{\cal G}_0\to\infty}\atop{y_0 = \sqrt{\pi/(2\alpha)}}}
  {\longrightarrow} \frac1{p_0-m},
\end{equation}
because if $\Psi$ is close to $1$ then the log term can be neglected
in \eqref{sol}, and we find $\Psi = 1+y_0 E$. 

For large values of $\Psi$, on the other hand, $y_0$ can be
neglected. Then the integral can be evaluated as
\begin{equation}
  \sqrt\pi\mathop{\mathrm{erfi}}(\sqrt{ \ln \Psi}) = E.
\end{equation}
For large $\Psi$ values it behaves as
\begin{equation}
  \frac{\Psi}{\sqrt{\ln\Psi}} = E,\qquad \mathrm{for\,large\,}E,\,\Psi.
\end{equation}

\subsection{Numerical solution}

Now let us turn to the numerical study of the system, based on
\cite{Jakovac} and \cite{Mati}: we determine the discontinuity of the
self-energy self-consistently. The discontinuity of \eqref{2PIsprep}
now reads
\begin{equation}
  \label{discSigma2pi}
  \Disc_{p_0} \Sigma(p) = \frac{\alpha}{\pi} \int\limits_0^{p_0} \!
  d\omega  (p_0-\omega) \rh(\omega).
\end{equation}
Knowing the discontinuity of the self-energy, we can use the
Kramers-Kronig relation to restore the complete self-energy:
\begin{equation}
  \label{KK}
  \Sigma(p) = \int\limits_{-\infty}^\infty \frac{d\omega}{2\pi}\,\frac{
    \Disc_{\omega}\,i\Sigma(\omega,\k)}{p_0-\omega+i\ep}.
\end{equation}
While \eqref{discSigma2pi} is a completely finite expression, in the
Kramers-Kronig relation we will find divergences. This corresponds to
the divergences of the self-energies which must be made finite by
applying the appropriate counterterms. Technically one can regularize
the integral in \eqref{KK} and then make it finite with counterterms,
or use the (twice) subtracted form of the Kramers Kronig relation. To
see how it works, we determine the one-loop result from the tree level
spectral function and the dimensional regularization of the
Kramers-Kronig equations (interpreting $\omega\to \sqrt{\omega^2}$):
\begin{equation}
  -2\alpha \mu^{2\ep}
  \pint{{1-2\ep}}\omega\,\omega\,(p_0-m-\omega)^{-1}
  =\frac{\alpha}{2\pi} (p_0-m)\left[\frac1\ep -2\ln\frac{m-p_0}{\mu} +
    \ln\pi+1\right].
\end{equation}
The divergence structure is the same, and also the \MSbar\ scheme
result is the same as in \eqref{Sigma1} (the different finite parts
are due to the different regularization method).

Now we can set up an algorithm to solve \eqref{discSigma2pi}. We
choose an arbitrary spectral function as a starting one (practically
the free spectral function), then follow the following steps:
\begin{description}
\item[\hspace*{1em}step 1:] compute the discontinuity of the self-energy using
  \eqref{discSigma2pi}
\item[\hspace*{1em}step 2:] compute the complete self-energy using the
  Kramers-Kronig relation \eqref{KK}
\item[\hspace*{1em}step 3:] renormalize the self-energy with local
  counterterms. To fix the counterterms we used on-mass-shell (OM)
  renormalization scheme, ie. the real part of the self-energy at the
  mass shell is zero and its derivative is also zero
  \begin{equation}
    \label{rencond}
    \Re\Sigma(p_0=m)=0,\qquad
    \frac{d\Re\Sigma(p_0)}{dp_0}\biggr|_{p_0=m}=0.
  \end{equation}
  We note here that releasing the first condition yields a mass shift,
  releasing the second condition yields a finite wave function
  renormalization. But in all renormalization schemes it will remain
  true that near the (renormalized) mass shell the propagator behaves
  as ${\cal G}(p_0\approx m) = \zeta/(p_0-m)$.
\item[\hspace*{1em}step 4:] construct the new spectral function from
  the discontinuity of the propagator knowing the real and imaginary
  part of the self-energy as
  \begin{equation}
    \rh(p) = \frac{2\Im\Sigma(p)}{(p_0-m-\Re\Sigma(p))^2 +
      (\Im\Sigma(p))^2}.
  \end{equation}
\item[\hspace*{1em}step 5:] continue with step 1 until the process
  converges.
\end{description}
Integrations in the above algorithm are performed numerically. This
strategy was applied successfully for the $\Phi^4$ model in
\cite{Jakovac}.

The direct application of this strategy, however, this times
fails. Numerically what we can observe is that the spectral function
becomes more and more shallow, and pointwise it goes to zero $\lim_n
\rh_n(p)=0$. In order to see a convergence, we had to use a
supplementary step in the iteration after step 4:
\begin{description}
\item[\hspace*{1em}step 4':] use a rescaling of the generated spectral
  function:
  \begin{equation}
    \rh(p) \to  A \rh(B\, p)
  \end{equation}
  with appropriate $A$ and $B$ which can ensure convergence.
\end{description}
The appropriate values can be found by inspection, but the actual
values are not too important (we used $A=73$ and $B=11$ in our
numerics). In this way finally we succeeded to see convergence in the
spectral function.

The numerical reason of this behavior is that the exact spectral
function has a discontinuity at the mass shell, and -- apart from this
single point -- it has always negative derivative. Numerically,
however, we cannot have a jump, since in all regularizations equation
\eqref{discSigma2pi} yields $\rh(p_0\approx m)\sim (p_0-m)^n$ where
$n\ge 2$. Since the exact curve starts to bend downwards, the
recursion tries to lower the spectral function in order to have
smaller derivative near the mass shell. Since the spectral function
has to be positive, these requirements can be satisfied only with
$\rh=0$. With the continuous rescaling we can achieve that the
numerically badly conditioned part, the vicinity of the mass shell,
becomes smaller and smaller.

\begin{figure}[htbp]
  \centering
  \includegraphics[width=8cm]{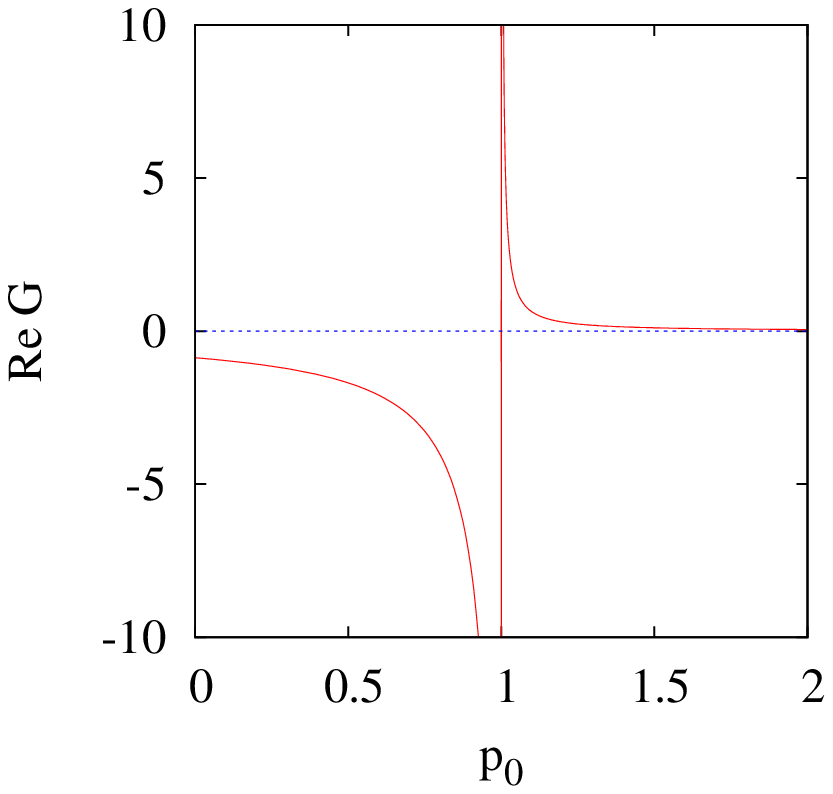}\qquad
  \includegraphics[width=8cm]{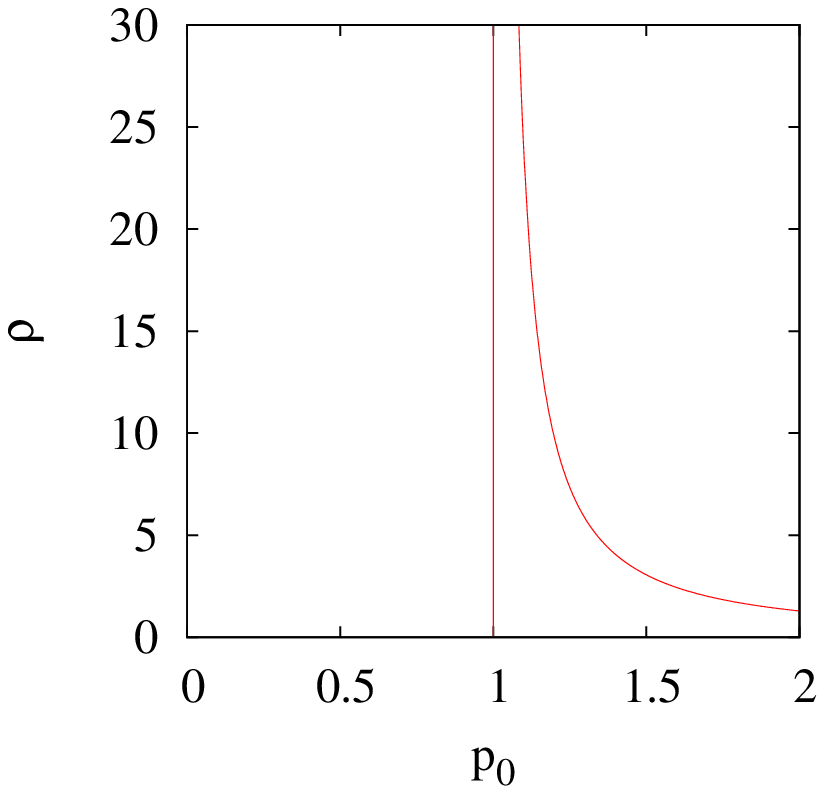}
  \caption{The real part and discontinuity of the 2PI propagator.}
  \label{fig:1}
\end{figure}

\begin{figure}[htbp]
  \centering
  \includegraphics[width=8cm]{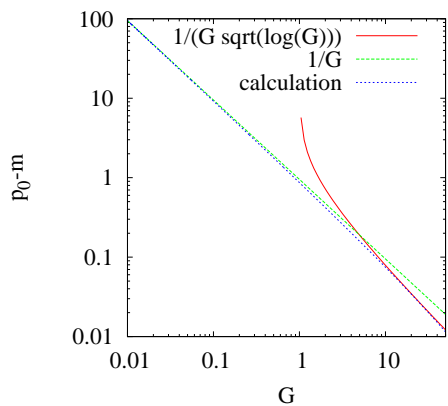}
  \caption{The expected asymptotics plotted on the data}
  \label{fig:2}
\end{figure}

The numerical results can be seen on Figure \ref{fig:1}.
The expected asymptotics can be nicely identified on the calculation
(cf. Fig. \ref{fig:2}).
This also proves implicitly that the strategy to resolve
the aforementioned numerical problem with the 2PI equation was correct.

If we compare the 1-loop, the 2PI and the exact results we see that
there is not too much improvement. The IR problem near the mass shell
which made the 1-loop calculation unreliable, \emph{seems} to be
cured, but in fact the result is not closer to the exact one as the
one-loop result. The physics of the deep infrared photons cannot be
described by the 2PI approximation.

\section{Schwinger-Dyson equations and Ward-identities}
\label{sec:SD}

The next level of the approximations is based on the Schwinger-Dyson
equations. For the Bloch-Nordsieck model in Feynman gauge it can be
written as
\begin{equation}
  \Sigma(p)=-ie^2\pint4k G(k) {\cal G}(p-k) u_\mu \Gamma^\mu(k;p-k,p),
\end{equation}
where $\Gamma^\mu$ is the vertex function.

For the vertex function there is another exact equation, coming from
the current conservation. This results in the Ward-identity analogous
to the QED  case \cite{PeshkinSchroeder}:
\begin{equation}
  k_\mu \Gamma^\mu(k;p-k,p) ={\cal G}^{-1}(p) - {\cal G}^{-1}(p-k).
\end{equation}

In this model, however, the vertex function is proportional to
$u^\mu$. In principle the Lorentz-index in this model can come from
$u^\mu$ or from any of the momenta. But, since the fermion propagator
depends on the 4-momentum in the form $u_\mu p^\mu$, the fermion-photon
vertex does not depend on the momentum components which are orthogonal to
$u_\mu$. Therefore the Lorentz-index which comes from $q^\mu$ in fact
comes from the longitudinal part of $q^\mu$, ie. proportional to
$u^\mu$. So we can write $\Gamma^\mu(k;p,q) = u^\mu
\Gamma(k;p,q)$. 

This gives us the possibility that from the Ward identities we
\emph{exactly} determine the vertex function. The Ward-identity for
the current conservation yields then in case when $u=(1,0,0,0)$:
\begin{equation}
  k_\mu \Gamma^\mu(k;p-k,p) = k_0 \Gamma(k;p-k,p) = {\cal G}^{-1}(p) -
  {\cal G}^{-1}(p-k) \tehat \Gamma(k;p-k,p) = \frac{{\cal G}^{-1}(p)-{\cal G}^{-1}(p-k)}{k_0}.
\end{equation}
Therefore we find
\begin{equation}
  \Sigma(p)=-ie^2\pint4k \frac{G(k)}{k_0} {\cal G}(p-k)
  \left({\cal G}^{-1}(p) - {\cal G}^{-1}(p-k)\right).
\end{equation}
This is an exact equation in the Bloch-Nordsieck model. Now we will
solve this equation in the renormalized theory, and demonstrate that
the solution is indeed identical with the Bolch-Nordsieck solution
presented in Section \ref{sec:1}.

In the second term ${\cal G}^{-1}(p-k)$ drops out, resulting in an integral
\begin{equation}
  -ie^2\pint4k \frac{G(k)}{k_0} = 0,
\end{equation}
because of $k_0\to-k_0$ symmetry. What remains is
\begin{equation}
  \label{DSWard}
   \Sigma(p)={\cal G}^{-1}(p) (-ie^2)\pint4k \frac{G(k)}{k_0} {\cal G}(p-k)  .
\end{equation}

This form is true in the original model, we shall now find the
renormalized form. First we adapt the wave function renormalization
for the fermionic fields which changes the bare propagator to
$1/(Z p_0 - (m+\delta m))$ where $Z=1+\delta Z$. We will assume that the
mass shell remains the same, then $m+\delta m =Z m$, and the free
propagator will be $1/(Z(p_0-m))$. We will use also the notation
$e_b=e+\delta e$. The full propagator then reads
\begin{equation}
  {\cal G}^{-1}(p) = Z(p_0-m) -\Sigma(p),
\end{equation}
Using \eqref{DSWard} we find the equation
\begin{equation}
  \label{Gex1}
  {\cal G}(p) = \frac{\zeta(p_0)}{p_0-m+i\ep}.
\end{equation}
where
\begin{equation}
  \label{zetadef}
  \zeta(p_0) = \frac{1+J(p_0)}Z\qquad\mathrm{and}\quad J(p_0) = -ie_b^2\pint4k
  \frac{G(k)}{k_0} {\cal G}(p-k).
\end{equation}
$\zeta(p_0)$ can be interpreted as a running wave function
renormalization constant.

With a spectral representation
\begin{equation}
  J(p_0) = \int\limits_0^\infty\!\frac{d\omega}{2\pi} \, \rh(\omega)\,
  I_1(\omega-p_0+i\ep),\qquad \mathrm{where}\quad I_1(a) =
  ie^2\pint4k \frac1{k_0^2-\k^2+i\ep}\, \frac1{k_0}\, \frac1{a+k_0}.
\end{equation}
In the Appendix we evaluate $I_1(p_0)$, and we find
\begin{equation}
  \label{Jrep1}
  J(p_0) = \frac{e_b^2}{4\pi^2}\int\limits_{-\infty}^\infty\!
  \frac{d\omega}{2\pi} \, \rh(\omega)\,  \left[{\cal D}_\ep -
  \ln\frac{\omega-p_0+i\ep}{\mu}\right].
\end{equation}
We rewrite it into \eqref{zetadef}, then, assuming normalizable spectral
function, after some algebraic manipulation we find
\begin{equation}
  \zeta(p_0) = \frac{\displaystyle \frac1{\alpha_b} + \frac1\pi{\cal
      D}_\ep - \frac1{\pi} \int\limits_{-\infty}^\infty\!
    \frac{d\omega}{2\pi}\, \rh(\omega)\,
    \ln\frac{\omega-p_0+i\ep}{\mu}}{Z/\alpha_b}.
\end{equation}
We may assume that the explicit integral is not UV divergent (it can be
checked a posteriori, or, as in the present case, knowing the exact
solution). Then the above equation can be made finite by requiring
\begin{equation}
  \frac1{\alpha_b} + \frac1\pi{\cal D}_\ep  = \frac1{\alpha_r},\qquad
  \frac{Z}{\alpha_b} =\frac{z_r}{\alpha_r}.
\end{equation}
where $\alpha_r$ and $z_r$ are finite. This form can be interpreted
physically as the appearance of the renormalized coupling $\alpha_r$
and the finite wave function renormalization $z_r$. We note that the
coupling constant renormalization equation agrees with the
nonperturbative  coupling constant renormalization in the O(N) models \cite{ONmodelren}.

Now we find
\begin{equation}
  \zeta(p_0) = \frac1{z_r}\left(1 - \frac{\alpha_r}{\pi}
    \int\limits_{-\infty}^\infty\! \frac{d\omega}{2\pi}\,
    \rh(\omega)\, \ln\frac{\omega-p_0-i\ep}{\mu}\right).
\end{equation}
This function depends on the arbitrary scale $\mu$, but the physics,
of course, must be $\mu$ independent. This can be achieved by
appropriately changing the $z_r$ and $\alpha_r$ constant when we
change $\mu$. The $\mu$-independence of $\zeta(p_0)$ requires (using
the sum rule for $\rh$):
\begin{equation}
  \frac{d\zeta(p_0)}{d\ln\mu} = -\frac1{z_r^2} \frac{dz_r}{d\ln\mu}
  \left(1 - \frac{\alpha_r}{\pi} \int\limits_{-\infty}^\infty\!
    \frac{d\omega}{2\pi}\, \rh(\omega)\,
    \ln\frac{\omega-p_0-i\ep}{\mu}\right) -
  \frac1{z_r\pi}\frac{d\alpha_r}{d\ln \mu}
  \int\limits_{-\infty}^\infty\! \frac{d\omega}{2\pi}\,  \rh(\omega)\,
  \ln\frac{\omega-p_0-i\ep}{\mu} + \frac1{z_r} \frac{\alpha_r}{\pi}=0.
\end{equation}
This can be satisfied if
\begin{equation}
  -\frac1{z_r^2} \frac{dz_r}{d\ln\mu}+\frac1{z_r}
  \frac{\alpha_r}{\pi}=0,\qquad \frac1{z_r^2}
  \frac{dz_r}{d\ln\mu}\frac{\alpha_r}{\pi} -
  \frac1{z_r\pi}\frac{d\alpha_r}{d\ln \mu} =0.
\end{equation}
The second equation means $z_r=\alpha_r/\alpha_0$ where $\alpha_0$ is
a constant; the first equation then reads
\begin{equation}
  \frac{d\ln z_r}{d\ln\mu} = \frac{\alpha_r}{\pi}\tehat
  \frac{d\alpha_r}{d\ln\mu} = \frac{\alpha_r^2}{\pi}  \tehat
  -\frac1{\alpha_r(\mu)} +\frac1{\alpha_r(\mu_0)}=\frac1\pi
  \ln\frac{\mu}{\mu_0}  \tehat
  \alpha_r(\mu) = \frac{\alpha_r(\mu_0)}{\displaystyle
    1+\frac{\alpha_r(\mu_0)}\pi \ln\frac{\mu_0}{\mu}}
\end{equation}

Using the normalizability of $\rh$ we finally find
\begin{equation}
  \zeta(p_0) = \frac{\alpha_0}{\pi}
  \int\limits_{-\infty}^\infty\! \frac{d\omega}{2\pi}\, \rh(\omega)\,
  \ln\frac\Lambda{\omega-p_0-i\ep},\qquad \Lambda = \mu e^{\frac\pi{\alpha_r}}.
\end{equation}
The $\alpha_0$ and the scale $\Lambda$ are renormalization group
independent quantities (ie. independent of the scale $\mu$), these
characterize the renormalization scheme. The appearance of a scale
$\Lambda$ is the manifestation of dimensional transmutation. Now,
instead of that scale $\Lambda$ it is worth to use $M$ for which
$\Re\zeta(M)=0$. Clearly $M\approx\Lambda$ if $\Lambda\gg m$. Then with
differentiating $\zeta$ with respect to $p_0$ we find
\begin{equation}
  \frac{d\zeta(p_0)}{dp_0} = -\frac{\alpha_0}{\pi} \int\limits_{-\infty}^\infty\!
  \frac{d\omega}{2\pi}\, \frac{\rh(\omega)}{p_0-\omega+i\ep} = -{\cal
    G}(p_0) \tehat \zeta(p_0)=\frac{\alpha_0}\pi
  \int\limits_{p_0}^M d\omega\, {\cal G}(\omega).
\end{equation}
This gives finally
\begin{equation}
  \label{Gex2}
  (p_0-m){\cal G}(p) =\frac{\alpha_0}{\pi} \int\limits_{p_0}^M\!
  d\omega\, {\cal G}(\omega).
\end{equation}
By differentiation with respect to $p_0$ we find
\begin{equation}
  (p_0-m){\cal G}' + {\cal G} = -\frac{\alpha_0}{\pi} {\cal G} \tehat
  {\cal G}(p) = g_0 (p_0-m)^{-1-\frac{\alpha_0}{\pi}},
\end{equation}
where $g_0$ is an arbitrary constant. This is indeed the solution of
Bloch and Nordsieck \eqref{Gex}, now in terms of the renormalized quantities.

But we also see that the condition ${\cal G}(p_0=M)=0$ can be
satisfied only with $g_0=0$. This is in close relation with the fact
that at the mass shell $p_0\approx m$, the propagator (and its
discontinuity) is not integrable.

The lesson of this analysis is that the deep IR physics is well
describable by the Schwinger-Dyson equation, truncated in a way which
respects the Ward-identities. As we have seen, this strategy is
renormalizable and exact in case of the Bloch-Nordsieck model.

A big advantage of this approach is that, besides being exact in the
IR, it can be easily generalized to other theories. So we expect that
in QED the Schwinger-Dyson equations truncated in the way we have done
it in the Bloch-Nordsieck theory will represent the exact result well
in the problematic deep IR regime.

\section{Conclusions}
\label{sec:conc}

In this paper we examined the exactly solvable Bloch-Nordsieck model
from the point of view of different perturbative methods. We first
reviewed the known method to obtain the exact solution
\cite{BogoljubovShirkov}. Then the different levels of approximations,
like the one-loop level perturbation theory, the 2PI resummation and
the truncated Schwinger-Dyson equations were studied. The 1-loop
result exhibits an IR sensitivity when we approach the mass shell
which renders the theory ill-defined. The self-energy (2PI)
resummation reorganizes the perturbative series in a way that this IR
problem disappears. This does not mean, however, that the result
itself would be closer to the exact one, only the explicit IR
sensitivity cannot be seen. On the other hand, the Schwinger-Dyson
equations, truncated in a way that the Ward-identities are satisfied
yield the exact result in the Bloch-Nordsieck model. This is a new way
of obtaining the exact solution in the Bloch-Nordsieck model. And,
while the original solution method is very hard to generalize to other
theories, the generalization of the specially truncated
Schwinger-Dyson equations is straightforward.

\begin{acknowledgments}
  The authors thank useful discussions with T.S. B\'\i r\'o, F. Csikor
  and A. Patk\'os. This work is supported by the Hungarian Research Fund
  (OTKA) under contract No. K68108.  
\end{acknowledgments}

\appendix
\section{Details of the one loop calculation}

The one-loop contribution to the self-energy reads, with a generic $u$
vector in Feynman gauge:
\begin{equation}
  \Sigma = -ie^2u^2 \pint4k \frac1{k^2+i\ep}\,\frac1{u^\mu
    (p_\mu-k_\mu)-m+i\ep}.
\end{equation}
This is Lorentz-invariant, if we a Lorentz transformation both on $u$
and $p$. So we may choose a special frame where $\Lambda
u=(u_0,0,0,0)$. If $u$ is a proper 4-velocity, then $u_0=1$; if it is
$u=(1,\mathbf{v})$, then $u_0=\sqrt{1-\mathbf{v}^2}$, but still
constant, since $\v$ is a parameter of the theory. We find then
\begin{equation}
  \Sigma = e^2u_0 I_0(\frac m{u_0}-p_0-i\ep),\qquad I_0(a) =
   i\pint4k \frac1{k^2+i\ep}\,\frac1{a+k_0}. 
\end{equation}
Thus it is enough to consider $I_0$ only. There we transform to
positive frequency integrals
\begin{equation}
  I_0(a) = i\pint4k \frac1{k^2+i\ep}\,\frac1{a+k_0} =
  \frac{ia}\pi \int\limits_0^\infty \!dk_0\,\pint3\k
  \frac1{k_0^2-\k^2+i\ep}\,\frac1{a^2-k_0^2} = \frac{a}\pi
  \int\limits_0^\infty \!dk_0\,\pint3\k \frac1{k_0^2+\k^2}\,\frac1{a^2+k_0^2},
\end{equation}
where in the last step we performed Wick rotation (the choice of the
imaginary part of $a$ is crucial for the direction of the rotation on
the complex plane).

Now we can write up the integral in $k_0$ and $\k$ space, in the
latter using $3-2\ep$ dimensions:
\begin{equation}
  I_0 = a\,\mu^{2\ep}\int\limits_0^\infty\frac{dk_0}{\pi}
  \pint{{3-2\ep}}\k \frac1{k_0^2+\k^2} \frac1{a^2+k_0^2}.
\end{equation}
We use the relation
\begin{equation}
  \mu^{2\ep}\pint{{d-2\ep}}k f(k^2) =
  \frac{2(4\pi\mu^2)^\ep}{(4\pi)^{d/2}\Gamma(d/2-\ep)}
  \int\limits_0^\infty\!dk\,k^{d-1+2\ep}f(k^2) =
  \frac{(4\pi\mu^2)^\ep}{(4\pi)^{d/2}\Gamma(d/2-\ep)}
  \int\limits_0^\infty\!dz\,z^{\frac d2-1-\ep}f(z)
\end{equation}
to proceed as
\begin{eqnarray}
  I_0 &&= \frac{a}\pi \int\limits_0^\infty\!dk_0\,\frac1{a^2+k_0^2}
  \frac{(4\pi\mu^2)^\ep}{(4\pi)^{3/2} \Gamma(\frac32-\ep)}
  \int\limits_0^\infty\!dz\,z^{\frac 32-1-\ep}(k_0^2+z)^{-1}
  = \frac{a(4\pi\mu^2)^\ep \Gamma(-\frac12+\ep)}{8\pi^2\sqrt\pi}
  \int\limits_0^\infty\!dk_0\,\frac{k_0^{1-2\ep}}{a^2+k_0^2} =\nn&&=
  \frac{a \Gamma(-\frac12+\ep)\Gamma(1-\ep)}{16\pi^2\sqrt\pi}
  \left(\frac{4\pi\mu^2}{a^2}\right)^\ep \Gamma(\ep) =
  \frac{-a}{8\pi^2} \left[\frac1\ep - 2\ln\frac{a}{\mu}
    +2+\ln\pi-\gamma_E\right].
\end{eqnarray}
We write it as
\begin{equation}
  I_0 = \frac{-a}{4\pi^2} \left[{\cal D}_\ep -
    \ln\frac{a}{\mu}\right],
\end{equation}
where
\begin{equation}
  {\cal D}_\ep = \frac1{2\ep} +1+\frac{\ln\pi-\gamma_E}2.
\end{equation}
Therefore
\begin{equation}
  \Sigma = (u_0p_0-m)\frac{e^2}{8\pi^2} \left[\frac1\ep -
    2\ln\frac{u_0p_0-m}{u_0\mu} +2+\ln\pi-\gamma_E\right].
\end{equation}

We also need to compute
\begin{equation}
  I_1(a) = i\pint4k \frac1{k^2+i\ep}\,\frac1{k_0}\,\frac1{a+k_0} =
  \frac{-i}\pi \int\limits_0^\infty \!dk_0\,\pint3\k
  \frac1{k_0^2-\k^2+i\ep}\,\frac1{a^2-k_0^2} = -\frac1a I_0(a) =
  \frac1{4\pi^2} \left[{\cal D}_\ep - \ln\frac{a}{\mu}\right].
\end{equation}

\end{document}